# Stimulated Brillouin Scattering Thresholds in Optical Fibers for Lasers Linewidth Broadened with Noise

**V. R. Supradeepa**

*OFS Laboratories, 19 Schoolhouse Road, suite 105, Somerset, NJ 08873, USA*
*supradeepa@ofsoptics.com*

**Abstract:** Phase and/or intensity modulation techniques to broaden the Linewidth of an optical source are well known methods to suppress stimulated Brillouin scattering (SBS) in optical fibers. A common technique used to achieve significant bandwidth enhancement in a simple fashion is to phase modulate with a filtered noise source. We will demonstrate here that, in this case the stochastic nature of noise requires an inclusion of length dependent corrections to the SBS threshold enhancement. This effect becomes particularly significant for short fiber lengths common to most high power fiber amplifiers.

## 1. Introduction

In recent years there has been significant interest in scaling the power of narrow linewidth fiber lasers [1] to the kW levels. One of the key motivators is possibility of combining multiple narrow linewidth lasers either through coherent combination or wavelength multiplexing to create very high power sources [2-4]. Stimulated Brillouin scattering (SBS) is the primary non-linearity affecting narrow linewidth fiber lasers and amplifiers and at these power levels, its management is of utmost importance. SBS causes backward scattered light at a slightly lower frequency to grow exponentially with the input signal power. It is often characterized in a system by a power threshold which is the power level at which the backward scattered light reaches a certain fraction of the signal light. Several threshold definitions exist depending on this ratio (3-dB, 20-dB, 30-dB etc) [5]. In this work, we will look primarily at the relative change in thresholds which is mostly independent of the actual definition used as long as it is consistent.

A well known strategy to suppress SBS is to broaden the Linewidth of the laser [6, 7]. The Brillouin process has a Lorentzian gain profile $g_B(f)$ characterized by a bandwidth $\Delta v_B$ which has a value in 10s of MHz for Silica fibers. This bandwidth is a measure of the response time of the process and is related to the acoustic phonon lifetime ($T_B$) by the expression $\Delta v_B = \Gamma_B / 2\pi = 1/2\pi T_B$ ($\Gamma_B$ is the gain bandwidth in angular frequency) [7]. Qualitatively, when the laser spectrum is much larger than $\Delta v_B$, we can look at it as being composed of multiple segments of width $\Delta v_B$. This corresponds to an enhancement in power threshold given by the number of segments which is $\sim \Delta f / \Delta v_B$ where $\Delta f$ is the bandwidth of the laser. In more detail, if $S(f)$ is the laser spectrum, the enhancement in threshold (which we will refer to as the enhancement factor ($EF$)) is given by

$$EF = \frac{\sigma(S(f) \otimes g_B(f))}{\sigma(g_B(f))} \tag{1}$$

Where, $\sigma()$ is a measure of the spectral width. This can be the full width at half maximum (FWHM) for smooth spectra like the Lorentzian or and something more appropriate for complicated spectra. In case of a Lorentzian laser lineshape, the above equation reduces to the simple relation

$$EF = 1 + \frac{\Delta f}{\Delta v_B} \tag{2}$$

From the above equation we see that a laser Linewidth of 5GHz corresponds to over 100X improvement in SBS threshold for a gain bandwidth of 50MHz. For the purposes of coherent or wavelength combining, GHz class laser linewidths seem to be sufficient [2-4].

Linewidth broadening can be achieved either through directly modulating the laser diode or through external modulation of a single frequency seed. The common techniques include creating a frequency chirp, phase modulation with a sinusoid or phase modulation with noise waveforms. Driving with a sinusoid [6] generates discrete harmonics spaced by the repetition rate and by ensuring that the repetition rate of the sinusoid is larger than the gain bandwidth, an enhancement in threshold can be obtained. With phase modulation alone, the discrete harmonics are not uniform in intensity. This results in under utilization of the total spectral width. Another disadvantage with discrete spectra is that the threshold reduction is related to the number of lines rather than the total bandwidth. This necessitates the repetition rate to be close to the SBS gain bandwidth for spectral efficiency. To achieve a 5GHz bandwidth for example, the total number of lines need to be > 100 (at 50MHz gain bandwidth) which is quite difficult [8]. Chirped waveforms are another common modulation scheme [9]. For high power amplifiers, the interaction lengths are usually of the order of a few meters (corresponding to time windows of 10s of ns). So the chirp rate necessary to obtain GHz class bandwidths is $10^{17}$ Hz/s or greater which is again quite difficult to achieve in practice [10].

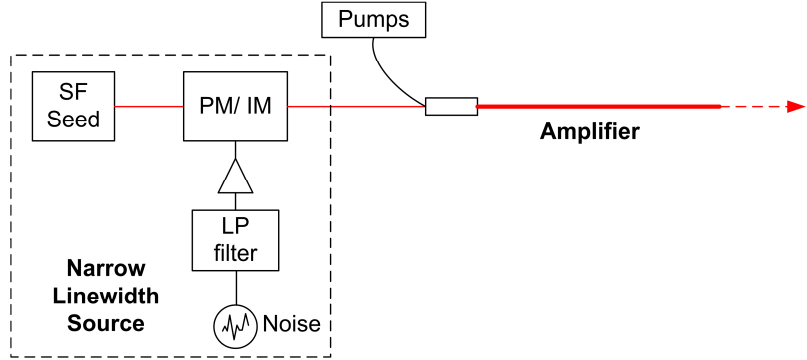

Fig. 1 Schematic showing a narrow linewidth input source to a power amplifier. The narrow linewidth source is generated by intensity (IM) and/or phase modulation (PM) with noise of a Single frequency (SF) seed. A low pass filter (LP) is used to control the bandwidth.

Due to the above reasons, broadening with noise is the commonly preferred strategy. Another key advantage is its inherent simplicity. Fig. 1 shows the schematic of a single stage high power amplifier with a narrow Linewidth seed source created by phase and/or intensity modulation with noise of a single frequency seed source. An amplifier is used to obtain suitable voltage levels and a low pass filter (possibly with further spectral shaping) allows for control of the optical bandwidth. Modulation with noise creates a continuum spectrum around the optical carrier with a lineshape depending on the filtering conditions [11]. Significant bandwidth enhancement can be achieved allowing for strong enhancement in SBS thresholds.

There has been anecdotal evidence that in high power fiber amplifiers, the SBS suppression achieved with noise modulation is smaller than what is anticipated. Recently, there was an interesting report from Zeringue, Dajani et al [12] where they investigate this effect in passive optical fibers through first principles numerical simulations. By numerically solving the equations for the optical field and the acoustic field together they demonstrated that the SBS threshold enhancement for short fiber lengths can be significantly smaller than what is expected from its bandwidth. This reduction was attributed to contributions from phase mismatched terms in the SBS process. For signal bandwidths in the GHz class, the coherence length of the signal is significantly smaller than the length of the fiber medium. In this case, the contributions from phase mismatched terms are expected to be small [6].

Here we offer a different perspective to the problem. We attribute this reduction to the stochastic nature of noise requiring a modified interpretation of the spectral width. We show that by incorporating the stochastic nature of noise, the observed enhancement of SBS in short optical fibers can be explained through the conventional model (Eq. 1). With a standard model for the modulating noise, namely a white Gaussian process, we derive an analytic expression for the length dependent reduction in enhancement.

## 2. Model

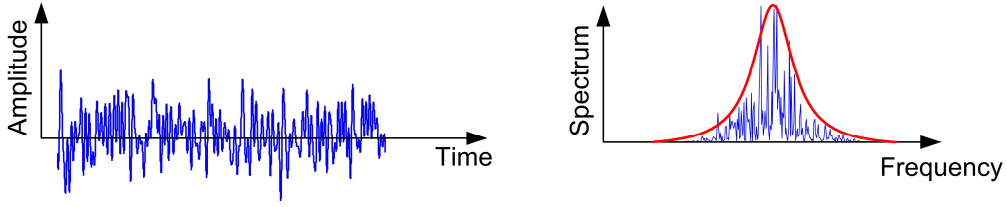

Fig. 2 An instance of the filtered noise time domain waveform used to modulate the phase of the optical carrier. Actual power spectrum and the power spectral density.

Figure 2 shows a specific instance of the low pass filtered noise waveform. This waveform is used to modulate the phase of an optical carrier and the spectrum obtained (blue) is shown. Also shown is the power spectral density (red) of the process (defined as the Fourier transform of the noise autocorrelation function). We see that the specific power spectrum is very different from the power spectral density. This is the case for random processes like noise while for deterministic waveforms, they are identical. Power spectral density is the appropriate measure for noise waveforms since the spectrum is a statistical quantity which changes every instance. The actual spectrum is of a broken nature and an appropriate measure of its spectral width is expected to be smaller than that of the power spectral density. If $x_k(t)$ is an instance of the waveform with power spectral density $\overline{S}(f)$, $X_k(f)$ is its transform and has its power spectrum defined by $S_k(f) = |X_k(f)|^2$ then [13, 14]

$$\langle S_k(f) \rangle = \lim_{k \to \infty} \frac{1}{k} \sum_k S_k(f) = \overline{S}(f) \tag{3}$$

This interesting detail explains the connection between the power spectral density and the instantaneous spectra for stochastic waveforms like noise. Different methods for measuring optical spectra like the optical spectrum analyzer, Fabry-perot interferometers, delayed heterodyne/homodyne methods etc all involve averaging and provide the power spectral density. Averaging inherent to the system (like the integration time of a photo-detector) result in the displayed spectrum to be the ensemble average of the different instances which is the power spectral density. Here we assume an Ergodic process whose time average is the same as the ensemble average. To obtain the SBS threshold enhancement, the relevant quantity is the width of the actual power spectrum. In the absence of sufficient averaging (which we will show is related to shorter fiber lengths) the width of the power spectrum is narrower than that of the power spectral density and this manifests as an apparent reduction in the enhancement factor in comparison to what is calculated using the measured linewidth (i.e. width of the power spectral density).

The SBS gain bandwidth $\Delta v_B$ gives us an intrinsic time scale. This also provides an equivalent length scale $L_B$ defined as the length travelled by light in the fiber in a time window of size $(1/\Delta v_B)$ and given by

$$L_B = \frac{c}{n_{fiber} \Delta v_B} \tag{4}$$

Where '$n_{fiber}$' is the effective refractive index of the optical fiber. We will model the SBS process as acting upon the signal in time windows of size $(1/\Delta v_B)$ or length segments of $L_B$. For a single segment, the effective power spectrum is just the power spectrum of the signal in that segment. For a length $L$ of the fiber, the number of instances now becomes $k = L/L_B$ and the effective spectrum would be the average of '$k$' instances. Fig. 3 shows the schematic of the model. The effective spectrum for a given length is the mean power spectrum of every segment preceding it. We see that length of the fiber plays the role of a smoothening parameter for the effective spectrum. For long fiber lengths, the number of instances will be large and from Eq. 3, the effective spectrum will be the same as the power spectral density. However, for short lengths, more complicated behavior is observed.

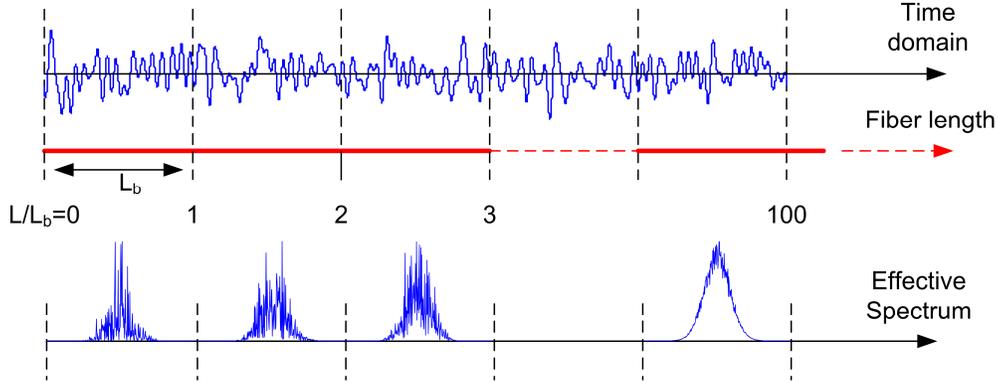

Fig. 3 The fiber (and the time domain waveform) used for intensity or phase modulation is considered in $L_B$ sized segments. The effective spectrum for a specific fiber length is the mean of the power spectrum of every segment preceding it. The effective spectrum smoothens out as the fiber length increases.

It is interesting to look at the physical basis for this model. The spectrally dependent gain for the SBS process is an exponential function of the signal spectrum. For a spectrally flat optical noise seeding the process, the Stokes light will acquire a spectral shape corresponding to an exponential of spectrum (scaled appropriately) of the signal light. Accounting for the time window inherent to the SBS process (due to its response time), the net optical gain for a system composed of multiple segments (windows) is the product of gain in individual segments. Due to exponential nature of gain, this will correspond to the exponential of the sum of spectra from individual segments. On normalizing for the total power, the effective spectrum is just the mean of the spectra from individual segments. In case of constant power propagating through the fiber, the averaging is standard while in the case of power varying with position; it will be a weighted average.

The analysis strategy will be to first obtain the statistics of the effective spectrum for a given fiber length. We will then obtain the enhancement factor from Eq. 1 together with a more appropriate measure for spectral width.

## 3. Analysis: passive, low-loss fibers

Let $x(t)$ represent the time domain envelope and $W_k(t)$ represent the windowing function for the '$k$'th segment. The effective signal for the segment is $x(t)W_k(t)$. The windowing function in a simplistic case can be a rectangular window of width given by $1/\Delta\upsilon_B$. More generally there can be additional structure arising from the Lorentzian lineshape of the SBS gain. Let $X_k^r(f)$ and $X_k^c(f)$ be the real and complex components of its Fourier transform. By multiplying with a window function in time, we have introduced spectral correlations (smoothening) in scales of the Brillouin gain bandwidth. So, it is sufficient to focus on the spectral samples $X_k^r(f_n)$ and $X_k^c(f_n)$ where $f_n = n\Delta\upsilon_B, n \in \mathbb{Z}$. This windowing captures in a way the convolution present in Eq. 1 and inversely, the convolution itself can be seen as an effect due to an inherent time window of the Brillouin response.

For phase or intensity modulation with a noise source which is a stationary Gaussian random process with zero mean, we have –

**Theorem 1.** $X_k^r(f_n)$, $X_k^c(f_n)$ are independent, identically distributed (iid) Gaussian random variables with zero mean.

In demonstrating this we will use the following established result without proof for linear functionals of Gaussian random processes [13, 14]

"If $x(t)$ is a zero mean Gaussian random process, $g_i(t), g_j(t)$ are real functions in $L^2$ with identical norms and are elements of an orthogonal set, then the linear functionals defined by $Z_i = \int_{-\infty}^{\infty} x(t)g_i(t)dt$ and $Z_j = \int_{-\infty}^{\infty} x(t)g_j(t)dt$ are independent and identically distributed Gaussian random variables."

For the case of amplitude modulation with a zero mean Gaussian process $a(t)$, we have

$$X_k^r(f_n) = \int_{-\infty}^{\infty} a(t)\overbrace{\cos(2\pi f_n t)W_k(t)}^{g_i(t)} dt$$

$$X_k^c(f_n) = \int_{-\infty}^{\infty} a(t)\overbrace{\sin(2\pi f_n t)W_k(t)}^{g_j(t)} dt \quad (5)$$

The independence of $X_k^r(f_n)$ and $X_k^c(f_n)$ is tested by the orthogonality integral

$$\frac{1}{2}\int_{-\infty}^{\infty} \sin(2*2\pi f_n t)|W_k(t)|^2 dt \quad (6)$$

For the model of a rectangular window $W_k(t)$ of width $(1/\Delta v_B)$, the orthogonality and identical norms of $\cos(2\pi f_n t)W_k(t)$ and $\sin(2\pi f_n t)W_k(t)$ is direct since the time window is an integral multiple of the period of the sinusoids. The theorem follows immediately from an application of the above result. In a more general situation, even with additional amplitude structure in the windowing function, Eq. 6 shows that, with an 'odd' or 'even' symmetry for $W_k(t)$, the orthogonality condition will still hold. For example, the Fourier transform of a Lorentzian gain profile has the form $\exp(-\alpha|t|)$ which has 'even' symmetry.

For the case of phase modulation with a zero mean Gaussian random process $\phi(t)$, the Fourier transform components are -

$$X_k^r(f_n) = \int_{-\infty}^{\infty} \cos\phi(t)\cos(2\pi f_n t)W_k(t)dt - \int_{-\infty}^{\infty} \sin\phi(t)\sin(2\pi f_n t)W_k(t)dt$$

$$X_k^c(f_n) = \int_{-\infty}^{\infty} \cos\phi(t)\sin(2\pi f_n t)W_k(t)dt + \int_{-\infty}^{\infty} \sin\phi(t)\cos(2\pi f_n t)W_k(t)dt \quad (7)$$

For a zero mean Gaussian process $\phi(t)$, $\cos\phi(t)$ and $\sin\phi(t)$ are both zero mean iid Gaussian random processes. Looking at each of the four terms individually in the above equations and using similar arguments as in the amplitude modulation case, it is clear that they are all mutually independent and identically distributed Gaussian variables. The random variables corresponding to the sum and difference of zero mean iid Gaussian random variables are both zero mean, iid Gaussian variables and hence the theorem follows. The above analysis can be further generalized to the case of simultaneous intensity and phase modulation with two independent noise waveforms.

***Theorem 2.*** The effective spectrum for $k$ instances $S_k(f_n)$ is "Chi-squared" distributed of order $2k$ [13]

We have

$$S_k(f_n) = \sum_{i=1}^{k}\left|X_i^r(f_n)\right|^2 + \sum_{i=1}^{k}\left|X_i^c(f_n)\right|^2 \quad (8)$$

"Chi-squared" is the distribution arising from the sum of squares of independent and identically distributed Gaussian random variables. Theorem 1 shows the two components of the Fourier transform within a time segment to be iid. Assuming that the noise waveform has a bandwidth much wider than $\Delta v_B$ (correlation time much shorter than the time segment), Fourier transform components arising from two non intersecting time windows are independent. The identical distribution arises from the stationarity of the noise process. So, the above equation has $2k$ independent and identically distributed terms which are squares of Gaussian random variables.

Equation 8 provides us the statistics for the effective power spectrum. For a measure of spectral width, we will use the following definition –

$$\sigma = \left\langle \frac{(\int S_k(f)df)^2}{\int S_k^2(f)df} \right\rangle \quad (9)$$

Over conventional measures of spectral width like the FWHM, the above definition captures much better the spectra which are broken. This definition is used frequently to obtain the spectral width of noise spectra and a similar definition is used to evaluate the effective area of optical fibers [7]. The spectral smoothening introduced by the windowing allows us to approximate well the above continuous definition by the following discrete sum -

$$\sigma = \Delta v_B \left\langle \frac{(\sum_n S_k(f_n))^2}{\sum_n S_k^2(f_n)} \right\rangle \quad (10)$$

Where $f_n = n\Delta v_B, n \in Z$. In the above definition, since the individual spectral samples are now independent of each other, the sum and expectation operations can be interchanged. The expression still has a quotient which cannot be readily simplified. We will use a simple physical argument to overcome this. The numerator is just a measure of the total energy in one segment squared. For CW operation, particularly in phase modulated cases, the variation in total power is low in time scales of the SBS process. Hence the above expression simplifies to

$$\sigma = \Delta v_B \frac{(\sum_n \bar{S}(f_n))^2}{\left\langle \sum_n S_k^2(f_n) \right\rangle} = \Delta v_B \frac{(\sum_n \bar{S}(f_n))^2}{\sum_n \left\langle S_k^2(f_n) \right\rangle} \quad (11)$$

Where $\bar{S}(f_n) = \langle S_k(f_n) \rangle$ and we have used the mutual independence of the spectral samples. For "chi-squared" distributions of order "$2k$" and mean $\bar{S}(f_n)$, we have

$$\left\langle S_k^2(f_n) \right\rangle = \frac{2k(2k+2)}{(2k)^2} \bar{S}^2(f_n) = \frac{k+1}{k} \bar{S}^2(f_n) \quad (12)$$

The effective spectral width is

$$\sigma = \frac{k}{k+1} \Delta v_B \frac{(\sum_n \bar{S}(f_n))^2}{\sum_n \left\langle \bar{S}^2(f_n) \right\rangle} = \frac{k}{k+1} \sigma_{ideal} \quad (13)$$

Where $\sigma_{ideal}$ is the spectral width corresponding to the power spectral density of the waveform. Converting these to enhancement factors using Eq. 1, we have a simple relation -

$$EF = \frac{k}{k+1} EF_{ideal} \quad (14)$$

Where $EF_{ideal}$ is the anticipated enhancement factor from the power spectral density alone. Substituting for k in terms of $L$ and $L_B$, we have

$$EF = \frac{(L/L_B)}{(L/L_B)+1} EF_{ideal} \quad (15)$$

Or in terms of fiber parameters

$$EF = \frac{\Delta v_B}{\Delta v_B + \dfrac{c}{n_{fiber} L}} EF_{ideal} \quad (16)$$

From the above equations, we see that the stochastic nature of noise has introduced a length dependent reduction of the enhancement factor compared to what is expected from the power spectral density. It is interesting to note that the reduction in enhancement depends only on the length of the fiber and does not depend on the linewidth. Furthermore it is independent of the modulation scheme (PM or IM) and spectral shaping of the noise waveform.

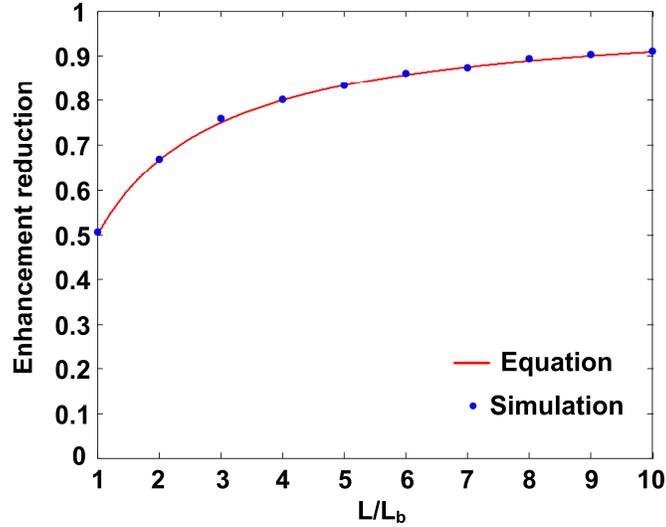

Fig. 4  Plot showing enhancement reduction (relative to ideal value) versus normalized length of the fiber.

Figure 4 shows the plot of enhancement reduction as a function of normalized length. Also shown is a simulation of our model (section 2) numerically. The two match very well indicating the correctness of the theoretical analysis of our model. For short lengths, the reduction in enhancement is significant. For silica fibers with a gain bandwidth of 50MHz, $L_B \sim 4m$. For a fiber length of 4m, the effective enhancement in threshold due to line broadening is only 0.5 times the ideal value. Even for fiber lengths as long as 40m (corresponding to a normalized length of 10), the actual enhancement is 10% smaller than the ideal value. This shows that though the reduction is most significant at short fiber lengths, even at much longer lengths, its effect can be substantial.

An experimental investigation is necessary to verify the correctness of this analysis. We do however have for comparison the results from the first principles numerical work reported in ref [12]. Fig. 5 shows the comparison between the two. Though the two do not match exactly, a good agreement between them is observed. This further strengthens our interpretation of the observed behavior. A point to note is that, in our model, we discretized the length parameter based on physical arguments and after analysis replaced the obtained discrete correction factor with its continuous counterpart. For short normalized lengths, particularly below 1 (smaller than the discretization) this can be a potential source of error. A continuous domain reformulation and a more through investigation of the SBS process for very short fibers would shed more light on this.

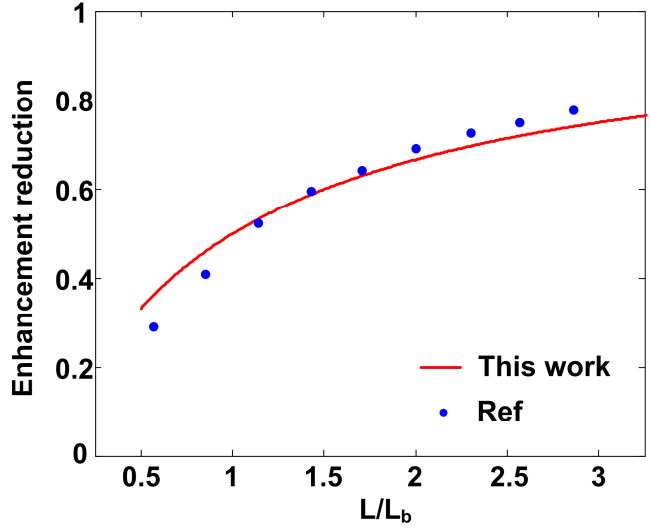

Fig. 5 Comparison between this work and the numerical results from ref [12]

## 4. Analysis: general case

In the previous section we looked at passive, low loss fibers. A length dependent correction factor was obtained for the SBS threshold which becomes significant at short lengths. Practically, such a situation often arises in fiber amplifiers where there is also a length dependent power variation. In such cases, the effective length cannot be identified with the physical length. In this section we will account for position dependent power variation and obtain a modified factor. We will then generalize this to systems having position dependent power as well as other fiber parameters. As before, we utilize a discrete model to obtain the correction factor and then generalize it to its continuous counterpart.

In our model, we account for the power variation with length by using a weighting parameter in the sum for effective spectra. Eq. 8 is modified as

$$S_k(f_n) = \sum_{i=1}^{k} w_i \left| X_i^r(f_n) \right|^2 + \sum_{i=1}^{k} w_i \left| X_i^c(f_n) \right|^2 \qquad (17)$$

Where $w_i$ is the weighting parameter. Since the effective contribution of each spectrum is proportional to the energy in each segment, we have

$$w_i = \frac{E_i}{\sum_{i=1}^{k} E_i} = \frac{P_i}{\sum_{i=1}^{k} P_i} \qquad (18)$$

$E_i$ and $P_i$ are the energy and mean power of each segment. The weighted sum for the effective spectra (Eq. 17) has a distribution which is referred to as the "Generalized chi-square distribution". Rewriting Eq. 17 in spectral components we have,

$$S_k(f_n) = \sum_{i=1}^{k} w_i S_2(f_n) \qquad (19)$$

$S_2(f_n)$ is "chi-squared" of order 2 with mean $\bar{S}(f_n)$, $\text{var}(S_2(f_n)) = \bar{S}^2(f_n)$, hence

$$\left\langle S_k^2(f_n) \right\rangle = (1 + \sum_{i=1}^{k} w_i^2) \bar{S}^2(f_n) \qquad (20)$$

We have,

$$\sum_{i=1}^{k} w_i^2 = \frac{\sum_{i=1}^{k} P_i^2}{(\sum_{i=1}^{k} P_i)^2} \qquad (21)$$

Interestingly, this form is similar to Eq. 10 and using the length discretization of $L_B$, the corresponding continuous form is readily identifiable as $L_B / L_{eff}$, where the effective length $L_{eff}$ is

$$L_{eff} = \frac{(\int P(z) dz)^2}{\int P(z)^2 dz} \qquad (22)$$

Using this in Eq. 13, we have

$$\sigma = \frac{L_{eff} / L_B}{1 + L_{eff} / L_B} \sigma_{ideal} \qquad (23)$$

In case of passive fibers, the reduction factor in spectral width will directly manifest as the reduction in enhancement factor. However in the case of fiber amplifiers, the backward SBS wave has both the non-linear gain component due to SBS and the rare-earth gain component. Modification of enhancement factor depends on their relative contributions. We can only say that the non-linear gain component due to SBS, $\exp(G_{SBS}(ideal))$ will be modified to $\exp(\frac{1 + L_{eff} / L_B}{L_{eff} / L_B} G_{SBS}(ideal))$. However, in high power fiber amplifiers, the rare-earth gain is usually much smaller than the non-linear gain when the SBS threshold occurs. In such cases, the effective enhancement factor can be approximated to

$$EF = \frac{L_{eff} / L_B}{1 + L_{eff} / L_B} EF_{ideal} = \frac{\Delta v_B}{\Delta v_B + \frac{c}{n_{fiber} L_{eff}}} EF_{ideal} \qquad (24)$$

In the above analysis we only considered a variation of power with length. In practice there can be circumstances where fiber parameters like the effective area, Brillouin gain co-efficient etc can change within the system. A simple example is when a larger delivery fiber is spliced to the gain fiber in a high power fiber amplifier. A minor modification to the above analysis is sufficient. In Eq. 18 we assumed the weighting parameter is dependent only on the mean power. The SBS gain however has the following form $\exp(\frac{g_B P}{A_{eff}} L)$ where $g_B$ is the Brillouin gain coefficient and $A_{eff}$ is the effective area. So, in cases where the other parameters are changing as well, the weighting parameter should contain the net factor $\frac{g_B P}{A_{eff}}$.

By generalizing the definition of $L_{eff}$ to

$$L_{eff} = \frac{(\int \frac{g_B(z)P(z)}{A_{eff}(z)} dz)^2}{\int (\frac{g_B(z)P(z)}{A_{eff}(z)})^2 dz} \quad (25)$$

We believe the same expressions for the spectral width reduction and enhancement reduction can be used.

## 5. Summary and discussion

Linewidth broadening of lasers with noise is a simple and common technique used to enhance the threshold for SBS. We showed that due to stochastic nature of noise, the obtained enhancement of SBS threshold is smaller than what is ideally expected from the power spectral density of the optical waveform. This effect becomes particularly important for shorter fibers. In this work, using a common model for noise, namely a Gaussian process, we obtained simple expressions for the reduction in enhancement factors as a function of length. We analyzed both cases where the power level and fiber parameters are constant or varying over the length of the fiber. We also looked at both phase and intensity modulation and demonstrated that the enhancement reduction factor is independent of the modulation scheme. We believe the model used here can be useful whenever there is a time dependent modification of the signal spectrum.

The statistical behavior of the noise broadened spectra has another interesting aspect. The net SBS gain will be a statistical variable with enhanced variation as the fiber lengths get shorter. This implies that the backward SBS signal can have significant intensity noise even when the forward signal has very low intensity noise. In particular, rare but possibly catastrophic events caused due to high gain events can occur. It is expected that when the mean gain is close to ideal gain, the variation in actual gain is small. When that is not the case, even though the mean gain might still have the system below SBS threshold, the variation in actual gain has the possibility to cause isolated but possibly dangerous events.

I would like to thank Chinmay Hegde, John Fini, Jeff Nicholson, Marc Mermelstein and the reviewers for helpful comments and discussions.